\documentclass[
  12pt,
  superscriptaddress,
  preprint,
  showpacs,
  nofootinbib,
  floatfix
]{revtex4-1}
\usepackage{hyperref}
\usepackage{graphicx}
\usepackage{subfigure}
\usepackage{amsmath}
\usepackage{amssymb}
\usepackage{slashed}
\usepackage{amsfonts}
\usepackage{xcolor}
\usepackage{multirow}
\usepackage{array}
\usepackage{mwe}
\usepackage{mathrsfs}
\usepackage{xr}

\begin{document}

\title{Purely Baryonic Weak Decays of Heavy Baryons in Skyrme Model}

\author{Chao-Qiang Geng}
\email{cqgeng@ucas.ac.cn}
\affiliation{School of Fundamental Physics and Mathematical Sciences,
Hangzhou Institute for Advanced Study, UCAS, Hangzhou 310024, China}

\author{Chao Han}
\email{hbarc@ucas.ac.cn}
\affiliation{School of Fundamental Physics and Mathematical Sciences,
Hangzhou Institute for Advanced Study, UCAS, Hangzhou 310024, China}

\begin{abstract}
Purely baryonic weak decays of heavy baryons are investigated within the framework of the Skyrme model. 
These decays belong to a new class of unobserved decay channels, which would help us to test the standard model,
particularly potential sources of CP violation in the baryonic sector.
By interpreting the heavy baryon as a bound state of a heavy meson and a baryon (Skyrmion), a direct calculation of the decay process $\Lambda_b \to p\,\bar p\,n$ is performed.
The resulting branching fraction is of $\mathcal{O}(10^{-6})$, in agreement with previous estimates.

\end{abstract}

\maketitle



\clearpage

\section{introduction}
The search for physics beyond the Standard Model (BSM) is one of the central goals of contemporary particle physics. Among the various possible signatures, new sources of CP violation are of particular interest, as they may help to explain the baryon asymmetry of the Universe and other fundamental open problems. While most experimental and theoretical studies have traditionally focused on mesonic decay channels, purely baryonic weak decays constitute a complementary and largely unexplored new class of processes with potential sensitivity to BSM interactions.
In this context, the decay of a heavy baryon into three light baryons provides a novel probe. Although such processes are conceptually simple, they capture essential features of baryonic weak dynamics and have only begun to receive attention in the literature~\cite{Geng:2016drz,Hsiao:2018dqg}. 
Since the decay modes involve four spin-1/2 baryons, T-odd triple-product correlations associated with the spins of the final state particles can be easily constructed.
They therefore represent a promising framework for future investigations of new physics.
Motivated by these considerations, the present work undertakes a detailed study of purely baryonic weak decays of heavy baryons within the Standard Model.

One of the classical approaches to heavy-baryon physics is provided by the Skyrme model within chiral perturbation theory (ChPT), which offers a low-energy effective description of QCD~\cite{Weinberg:1978kz,Weinberg:1990rz,Scherer:2002tk} and is primarily formulated to describe pion dynamics. To incorporate baryonic degrees of freedom, additional structures must be introduced.
Within this framework, baryons emerge as topological solitons (Skyrmions) of the chiral Lagrangian~\cite{Skyrme:1961vq,Skyrme:1962vh,Adkins:1983ya,Ma:2016npf}. This soliton picture has proven effective in describing various baryonic properties, including form factors~\cite{Braaten:1986md,Alberto:2015xea}.
When extended to incorporate heavy-flavor degrees of freedom~\cite{Wise:1992hn,Manohar:2000dt}, the Skyrme model further allows heavy baryons to be interpreted as bound states of a heavy meson and a soliton~\cite{Jenkins:1992se,Jenkins:1992zx,Weigel:2008zz}. This provides a unified framework for treating both light and heavy baryonic systems.

This paper presents a detailed theoretical study of the purely baryonic weak decay $\Lambda_b \to p\bar{p}n$. Within the bound-state picture of heavy baryons, the $\Lambda_b$ baryon is treated as a composite system consisting of a heavy meson bound to a soliton. This representation allows the decay rate to be computed in terms of the relevant hadronic matrix elements in a controlled and systematic manner.
In this context, time-like form factors associated with the weak current between nucleon states are required. To obtain these quantities, the method developed in Ref.~\cite{Alberto:2015xea} is adopted and extended. In particular, time-like form factors are constructed through analytic continuations of numerically determined space-like form factors, generalizing the original electromagnetic treatment to the weak-current case.
Using these ingredients within the Skyrmion framework, the decay rate can be derived and evaluated numerically. 

This paper is organized as follows. 
In Sec.~II, the basic building blocks within the framework of ChPT are introduced, including a brief overview of ChPT, the solitonic description of baryons, and the construction of baryon form factors in the Skyrme model. In particular, the time-like form factors relevant to the weak current are computed. The incorporation of heavy-flavor degrees of freedom and the interpretation of heavy baryons within the Skyrme model are then discussed.
Section~III presents a detailed analysis of the weak decay mechanism and the evaluation of the relevant matrix elements. Numerical results for the decay rates are given and discussed in Sec.~IV. Finally, Sec.~V summarizes the main results and discusses possible future extensions.

\section{Skyrme model}

This section introduces the interpretation of heavy baryons in the Skyrme model and presents the building blocks required for subsequent decay-rate calculations. Since several theoretical ingredients are involved, a collective overview is provided below. For further details, see Refs.~\cite{Adkins:1983ya,Holzwarth:1985rb,Jenkins:1992se,Weigel:2008zz,Ma:2016npf,Alberto:2015xea} and references therein.

\subsection{Pion}
In the framework of ChPT, quark and gluon degrees of freedom are integrated out via hadronization, leaving hadronic fields as the effective degrees of freedom. 
The basic building block of the chiral Lagrangian is the chiral field,  $U(x)=\displaystyle\exp\left(\frac{i\phi(x)}{F_\pi}\right),$
where \( \phi(x) \) denotes the pion fields and \( F_\pi \) is the pion decay constant. The leading-order chiral Lagrangian is given by $\mathcal{L}=\displaystyle\frac{F_{\pi}^2}{4}\mathrm{Tr}\left(\partial^\mu U\partial_\mu U^\dagger\right).$

\subsection{Baryons as Solitons in the Skyrme Model}

Within ChPT, baryonic degrees of freedom can be incorporated through the Skyrme model~\cite{Skyrme:1962vh,Adkins:1983ya}. In this approach, baryons emerge as topological soliton solutions, so-called Skyrmions, of the chiral field.
The Skyrme model extends the leading-order chiral Lagrangian by introducing a stabilizing higher-derivative term, known as the Skyrme term~\cite{Skyrme:1961vq,Skyrme:1962vh}. This term prevents soliton collapse and ensures the existence of finite-energy configurations. For static field configurations, the corresponding energy functional takes the form
\begin{equation}
\mathcal{L}_{\text{Skyr}}
=
\frac{F_\pi^2}{4}\,\mathrm{Tr}\!\left(\partial_\mu U\, \partial^\mu U^\dagger\right)
+\frac{1}{32 e^2}\,\mathrm{Tr}\!\left(
[U^\dagger \partial_\mu U,\, U^\dagger \partial_\nu U]\,
[U^\dagger \partial^\mu U,\, U^\dagger \partial^\nu U]
\right),
\end{equation}
where $e$ is the Skyrme coupling constant.
Minimization of this energy functional yields the classical Skyrmion solution. 
The Skyrmion configuration corresponds to a topologically nontrivial mapping from spatial infinity, compactified as $S^3$, to the group manifold $SU(2)$. The associated integer-valued topological charge is identified with the baryon number, establishing a direct connection between topology and baryon number conservation in low-energy QCD.

\subsubsection{Hedgehog Ansatz and Soliton Equation}
To obtain explicit finite-energy soliton solutions, a symmetric parametrization of the chiral field is required. Imposing spherical symmetry under combined spatial and isospin rotations naturally leads to the adoption of the hedgehog ansatz 
\begin{equation}
U(\mathbf{r}) = \exp\!\left[i\, \boldsymbol{\tau} \cdot \hat{\mathbf{r}}\, F(r)\right],
\end{equation}
where \( \boldsymbol{\tau} \) are the Pauli matrices, \( \hat{\mathbf{r}} = \mathbf{r}/r \), and \( F(r) \) is the radial profile function.

Substituting this ansatz into the Skyrme Lagrangian yields the classical equation of motion for \( F(r) \):
\begin{equation}
\left( \frac{1}{4} \tilde{r}^2 + 2 \sin^2 F \right) F'' 
+ \frac{1}{2} \tilde{r} F' 
+ \sin 2F \, (F')^2 
- \frac{1}{4} \sin 2F 
- \frac{\sin^2 F \sin 2F}{\tilde{r}^2} = 0,
\end{equation}
where \( \tilde{r} = e F_\pi r \). 
This nonlinear differential equation is solved numerically under the boundary conditions $F(0) = \pi,  F(\infty) = 0.$
The corresponding {soliton mass} is given by
\begin{equation}
M_{\text{Skyr}} = 4\pi \frac{F_{\pi}}{e} \int_0^\infty dr\, r^2 
\left\{
 \frac{1}{2} \left[ (F')^2 + 2 \frac{\sin^2 F}{r^2} \right]
+ \frac{1}{2} \frac{\sin^2 F}{r^2}
\left[ \frac{\sin^2 F}{r^2} + 2 (F')^2 \right]
\right\}.   
\end{equation}

\subsubsection{Collective Quantization of the Soliton}

The classical skyrmion solution carries topological charge but lacks definite quantum numbers such as spin and isospin. 
To obtain physical baryons, one introduces {collective coordinate quantization} by promoting the global SU(2) orientation to a time-dependent variable~\cite{Adkins:1983ya}:
\begin{equation}
\Sigma(\mathbf{x}, t) = A(t) \, \Sigma_0(\mathbf{x}) \, A^{-1}(t),
\end{equation}
where \( A(t) \in SU(2) \) and $\Sigma_0(\mathbf{x}) = \exp\!\left[i F(r)\, \hat{\mathbf{x}} \cdot \boldsymbol{\tau}\right].$
Quantization of the rotational degrees of freedom in \( A(t) \) leads to wavefunctions defined on the SU(2) collective space, whose eigenstates correspond to the nucleon and the \(\Delta\) resonance with definite spin and isospin quantum numbers. 

The baryon mass spectrum can be obtained through collective-coordinate quantization. The moment of inertia of the Skyrmion is given by 
\begin{equation}
    \mathcal{I}_{\text{Skyr}}= \frac{8 \pi}{3} \int_{0}^{\infty}  \, r^{2} dr \sin ^{2}F \left\{  F_{\pi}^{2} + \frac{1}{e^{2}} \left( (F')^{2}+ \frac{\sin ^{2}F}{r^{2}} \right) \right\}  ,
\end{equation}
and the nucleon mass takes the form $\displaystyle M_{N}= M_{\text{Skyr}} + \frac{3}{8 \mathcal{I}_{\text{skyr}}}$.

This quantized soliton framework allows systematic computation of baryonic observables such as magnetic moments, axial charges, and form factors~\cite{Adkins:1983ya,Braaten:1986md}. 
These results exhibit good agreement with experimental measurements, lending strong support to the solitonic interpretation.

\subsubsection{Electroweak Form Factors in the Skyrme Model}

The electroweak form factors of baryons can be systematically derived following the above mentioned collective quantization~\cite{Braaten:1986md}.
At leading order in \(1/N_c\), the electric and magnetic form factors \( G_E^{p/n}(q^2) \) and \( G_M^{p/n}(q^2) \), together with the axial and induced pseudoscalar form factors \( G_A(q^2) \) and \( G_P(q^2) \), obey the relations 
\begin{equation}
    \begin{aligned}
&G_E^{p}(q^{2})
= \frac{1}{2\Lambda_s}\!\left[t_1^{02}(-q^{2})-\frac{1}{3}t_2^{02}(-q^{2})\right]
+ \frac{1}{2}\, b_0^{00}(-q^{2}) , 
\\&
G_E^{n}(q^{2})
= -\frac{1}{2\Lambda_s}\!\left[t_1^{02}(-q^{2})-\frac{1}{3}t_2^{02}(-q^{2})\right]
+ \frac{1}{2}\, b_0^{00}(-q^{2}) ,
\\&
G_M^{p}(q^{2})
= \frac{M_N}{2\Lambda_s}\, b_1^{11}(-q^{2})
+ \frac{2M_N}{3}\, t_0^{11}(-q^{2}) ,
\,
G_M^{n}(q^{2})
= \frac{M_N}{2\Lambda_s}\, b_1^{11}(-q^{2})
- \frac{2M_N}{3}\, t_0^{11}(-q^{2}) ,
\\
&G_A(q^2)= -\frac{1}{9} (2 w_0^{00} (-q^2)-w_1^{00}(-q^2)) - \frac{1}{9}  q^2(w_0^{20} (-q^2)+w_1^{20}(-q^2)),
\\&
G_P(q^2)= \frac{4 M_N^2}{3} (w_0^{20}(-q^2)+ w_1^{20} (-q^2)),
    \end{aligned}
\end{equation}
where the $\Lambda_s$ is the moment of inertia of the soliton and the $b$, $t$ and $w$ functions are given by
\begin{equation}
    \begin{aligned}
t_ 0(r^2)&=t_ 1(r^2)=t_ 2(r^2)= \frac{1}{4} F_\pi^2 \
\frac{\sin^2F}{r^2}+ \frac{1}{e^2} \
\frac{\sin^2F}{r^2}[F^{'2}+\frac{\sin^2F}{r^2}],
\\
b_ 0(r^2)&= \frac{1}{2 \pi^2} \
\frac{\sin^2F}{r^2}(-F'),
\\
w_0(r^2) &= \frac{1}{2} F_\pi^2 \frac{\sin F}{r} \cos F 
+ \frac{2}{e^2} \frac{\sin F}{r} \cos F 
\left(F'^2 + \frac{\sin^2 F}{r^2}\right), 
\\
w_1(r^2) &= -(\frac{F_\pi^2}{2}+ \frac{4 sin^2F}{e^2 r^2}) F'.
\end{aligned}
\end{equation}
Their momentum-space representations are obtained through the spherical Bessel transform:
\begin{equation}
\label{expansion}
f^{mn}(q^2) = |q|^{-m} \int_0^\infty 4\pi r^2 \, dr \, 
j_m(|q|r) \, r^n f(r^2),
\end{equation}
where \( j_m(|q|r) \) are the spherical Bessel functions.

\subsubsection{time-like form factor}

At this stage space-like form factors are expressed in terms of the functions \( t \), \( b \), and \( w \), which are defined through the Skyrme profile function \( F(r) \). However, the matrix element \( \langle \bar p, n \lvert j^\mu \rvert 0 \rangle \) involves time-like momentum transfer and therefore requires the corresponding time-like form factors. This necessitates an analytic continuation from the space-like region to the time-like region.
A further complication arises from the fact that the Skyrme profile function \( F(r) \) is determined by a nonlinear differential equation. As a consequence, \( F(r) \) is known only through numerical solutions, and the resulting space-like form factors do not admit closed analytic expressions.

In this section, a multipoint Padé approximation of the form factors is employed to enable a controlled analytic continuation, following the approach of Ref.~\cite{Alberto:2015xea}. Within this framework, a systematic procedure is developed to construct timelike form factors from the corresponding spacelike ones. The method is then adopted to study the weak-current case in the present work. The corresponding $w$ functions are introduced to describe the weak-current form factors, whose explicit expressions will be given below.

To construct the Padé fitting formula, the asymptotic behaviors of the functions \( w \) are required as input constraints. These asymptotic forms restrict the admissible functional structure of the approximation and improve the numerical stability of the fit. They are obtained by adopting the standard pion tail for the hedgehog profile,
$F(r) \sim \frac{C}{r} e^{-M_\pi r}, \quad
\sin F \sim F, \quad
\cos F \sim 1, \quad
F'(r) \sim -M_\pi \frac{C}{r} e^{-M_\pi r}.$
For the function \( w_0 \), the leading contribution derives from
$\displaystyle\frac{\sin F}{r}\cos F \sim \frac{e^{-M_\pi r}}{r^2}.$
The next-to-leading term comes  from $\displaystyle\frac{\sin F}{r}\cos F\left(F'^2+\frac{\sin^2 F}{r^2}\right)$,
which behaves asymptotically as $\displaystyle\frac{e^{-M_\pi r}}{r^2} \frac{e^{-2M_\pi r}}{r^2}
=\frac{e^{-3M_\pi r}}{r^4}.$
Further contributions are more strongly suppressed, scaling as \( \sim e^{-3 M_\pi r} / r^6 \), and can be neglected at this order. Consequently, retaining the first two asymptotic terms is sufficient for the construction of the Padé approximant 
\begin{equation}
w_0(r)\sim A\frac{e^{-M_\pi r}}{r^2}+B\frac{e^{-3M_\pi r}}{r^4}.    
\end{equation}

For the function \( w_1 \), the leading contribution is given by 
$-F'\sim e^{-M_\pi r}/r ,$
while the next-to-leading term behaves as
$\displaystyle\sin^2F/r^2  (-F') \sim (\frac{e^{-2M_\pi r}}{r^4})(\frac{e^{-M_\pi r}}{r})=\frac{e^{-3M_\pi r}}{r^5}.$
Accordingly, the asymptotic form of \( w_1 \) is
$$w_1(r)\sim A\frac{e^{-M_\pi r}}{r}+B\frac{e^{-3M_\pi r}}{r^5}.$$

Based on these asymptotic behaviors, the fitting functions used in the Padé approximation can be parametrized as
\begin{equation}
    f_{\text{fit}}(r)=\sum_{a}e^{-\mu_a r}\frac{P^{(a)}{n_a}(r)}{Q^{(a)}{m_a}(r)},
\end{equation}
with $m_a=n_a+p_a$.

So for the $w$ functions,
\begin{itemize}
    \item $w_0$: two tails
  $((\mu_1,p_1)=(M_\pi,2)\to m_1=n_1+2)$,
  $((\mu_2,p_2)=(3M_\pi,4)\to m_2=n_2+4)$.
    \item $w_1$: two tails
  $((\mu_1,p_1)=(M_\pi,1)\to m_1=n_1+1)$,
  $((\mu_2,p_2)=(3M_\pi,5)\to m_2=n_2+5)$.
\end{itemize}

For the cases considered in this paper, the results can be further simplified using the Mittag-Leffler expansion due to the simple pole structure obtained from the Padé approximation. For each tail component:
\begin{equation}
 \frac{P_n(r)}{P_m(r)}
= \sum_{j=1}^{M}
\left[
\sum_{k=-\mu_j}^{-1} C_k^{(j)} (r - z_j)^k
\right],
\quad
\frac{P_n^{(a)}(r)}{Q_m^{(a)}(r)}=\sum_{j=1}^{m}\frac{R_{aj}}{r-z_{aj}},
\quad
R_{aj}=\frac{P_n^{(a)}(z_{aj})}{\left(Q_m^{(a)}\right)'(z_{aj})}.   
\end{equation}
Hence the analytic model becomes
\begin{equation}
    f_{\text{fit}}(r)=\sum_{a}\sum_{j=1}^{m_a}\frac{R_{aj}e^{-\mu_a r}}{r-z_{aj}}.
\end{equation}

Insert the fitted function \( f_{\text{fit}} \) into the definition in Eq.~\eqref{expansion}. The resulting expressions reduce to integrals of the generic form $\displaystyle\int_0^\infty dr \frac{r^k  e^{-(\mu \mp i q) r}}{r - z}.$
To evaluate these integrals, define the function
\begin{equation}
    H(\alpha,\beta):=\int_0^\infty \frac{e^{-\alpha r}}{r+\beta}dr
\quad (\text{Re}(\alpha)>0),
\qquad
H(\alpha,\beta)=e^{\alpha\beta}E_1(\alpha\beta).
\end{equation}
together with the corresponding moment kernels
\begin{equation}
    H_k(\alpha,-z):=\int_0^\infty \frac{r^k e^{-\alpha r}}{r-z} dr
=(-1)^k\frac{\partial^k}{\partial\alpha^k}H(\alpha,-z).
\end{equation}
These functions are closely related to the exponential integral and can be written explicitly in terms of \( E_1 \). Using $H(\alpha,-z)=e^{-\alpha z} E_1(-\alpha z),
$
the higher moments are obtained as $H_k(\alpha,-z)
= \displaystyle(-1)^k \frac{\partial^k}{\partial \alpha^k}
\left[ e^{-\alpha z} E_1(-\alpha z) \right].$

With these definitions, the specific building blocks appearing in the
\( w \) functions can be constructed straightforwardly.
\begin{itemize}
\item  \( w_0 \)-function building blocks

The function \( w_0 \) contains two asymptotic tails with $\mu_1 = M_\pi, \, \mu_2 = 3 M_\pi.$
For \( w_0^{00} \), the result is
\begin{equation}
\begin{aligned}
w_0^{00}(Q^2) &= \frac{2\pi}{iQ} \Bigg[ \sum_{j=1}^{m_1^{(0)}} R_{1j}^{(0)} \Big( H_1(\mu_1-iQ,-z_{1j}^{(0)}) - H_1(\mu_1+iQ,-z_{1j}^{(0)}) \Big) \\
&\quad + \sum_{j=1}^{m_2^{(0)}} R_{2j}^{(0)} \Big( H_1(\mu_2-iQ,-z_{2j}^{(0)}) - H_1(\mu_2+iQ,-z_{2j}^{(0)}) \Big) \Bigg].
\end{aligned}
\end{equation}
For $w_0^{20}$, one finds
\begin{equation}
\begin{aligned}
w_0^{20}(Q^2) &= 4\pi \Bigg[ \sum_{j=1}^{m_1^{(0)}} R_{1j}^{(0)} \Bigg( \frac{3}{2iQ^3} \Big[ H_{-1}(\mu_1-iQ,-z_{1j}^{(0)}) - H_{-1}(\mu_1+iQ,-z_{1j}^{(0)}) \Big] \\
&\quad -\frac{1}{2iQ} \Big[ H_1(\mu_1-iQ,-z_{1j}^{(0)}) - H_1(\mu_1+iQ,-z_{1j}^{(0)}) \Big] \\
&\quad -\frac{3}{2Q^2} \Big[ H_0(\mu_1-iQ,-z_{1j}^{(0)}) + H_0(\mu_1+iQ,-z_{1j}^{(0)}) \Big] \Bigg) \\
&\quad + \sum_{j=1}^{m_2^{(0)}} R_{2j}^{(0)} \Bigg( \frac{3}{2iQ^3} \Big[ H_{-1}(\mu_2-iQ,-z_{2j}^{(0)}) - H_{-1}(\mu_2+iQ,-z_{2j}^{(0)}) \Big] \\
&\quad -\frac{1}{2iQ} \Big[ H_1(\mu_2-iQ,-z_{2j}^{(0)}) - H_1(\mu_2+iQ,-z_{2j}^{(0)}) \Big] \\
&\quad -\frac{3}{2Q^2} \Big[ H_0(\mu_2-iQ,-z_{2j}^{(0)}) + H_0(\mu_2+iQ,-z_{2j}^{(0)}) \Big] \Bigg) \Bigg].
\end{aligned}
\end{equation}
\item  \( w_1 \)-function building blocks

Similarly, the function \( w_1 \) has two asymptotic tails, $\mu_1 = M_\pi, \, \mu_2 = 3 M_\pi.$
For \( w_1^{00} \), one finds
\begin{equation}
\begin{aligned}
w_1^{00}(Q^2) &= \frac{2\pi}{iQ} \Bigg[ \sum_{j=1}^{m_1^{(1)}} R_{1j}^{(1)} \Big( H_1(\mu_1-iQ,-z_{1j}^{(1)}) - H_1(\mu_1+iQ,-z_{1j}^{(1)}) \Big) \\
&\quad + \sum_{j=1}^{m_2^{(1)}} R_{2j}^{(1)} \Big( H_1(\mu_2-iQ,-z_{2j}^{(1)}) - H_1(\mu_2+iQ,-z_{2j}^{(1)}) \Big) \Bigg].
\end{aligned}
\end{equation}
For $w_1^{20}$, one has
\begin{equation}
\begin{aligned}
w_1^{20}(Q^2) &= 4\pi \Bigg[ \sum_{j=1}^{m_1^{(1)}} R_{1j}^{(1)} \Bigg( \frac{3}{2iQ^3} \Big[ H_{-1}(\mu_1-iQ,-z_{1j}^{(1)}) - H_{-1}(\mu_1+iQ,-z_{1j}^{(1)}) \Big] \\
&\quad -\frac{1}{2iQ} \Big[ H_1(\mu_1-iQ,-z_{1j}^{(1)}) - H_1(\mu_1+iQ,-z_{1j}^{(1)}) \Big] \\
&\quad -\frac{3}{2Q^2} \Big[ H_0(\mu_1-iQ,-z_{1j}^{(1)}) + H_0(\mu_1+iQ,-z_{1j}^{(1)}) \Big] \Bigg) \\
&\quad + \sum_{j=1}^{m_2^{(1)}} R_{2j}^{(1)} \Bigg( \frac{3}{2iQ^3} \Big[ H_{-1}(\mu_2-iQ,-z_{2j}^{(1)}) - H_{-1}(\mu_2+iQ,-z_{2j}^{(1)}) \Big] \\
&\quad -\frac{1}{2iQ} \Big[ H_1(\mu_2-iQ,-z_{2j}^{(1)}) - H_1(\mu_2+iQ,-z_{2j}^{(1)}) \Big] \\
&\quad -\frac{3}{2Q^2} \Big[ H_0(\mu_2-iQ,-z_{2j}^{(1)}) + H_0(\mu_2+iQ,-z_{2j}^{(1)}) \Big] \Bigg) \Bigg].
\end{aligned}
\end{equation}

\end{itemize}

As a result, the individual components of the functions \( w \) admit analytic representations in terms of pole residues and special functions. This formulation enables a controlled analytic continuation to the time-like region, implemented by the replacement \( Q \to i q \, (q>0) \), and yields the time-like form factors required for the decay-rate calculation.

\subsection{Heavy Meson}
When systems contain both light and heavy quarks, it is necessary to incorporate heavy-quark symmetry in addition to chiral symmetry. This requirement leads to the framework of heavy-meson ChPT~\cite{Wise:1992hn}, which systematically combines the low-energy chiral expansion with the heavy-quark effective theory (HQET) expansion~\cite{Manohar:2000dt}.
Within this framework, the heavy-meson field is conventionally defined as
\begin{equation}
H_a = \frac{1 + \!\not{v}}{2} \left( P_{a\mu}^{} \gamma^\mu - P_a \gamma_5 \right),
\end{equation}
where \( v^\mu \) is the four-velocity of the heavy meson, \( P_a \) and \( P_{a\mu} \) denote the pseudoscalar and vector heavy meson fields, respectively, and the subscript \( a \) labels the light flavor index.
This field transforms linearly under heavy-quark symmetry and nonlinearly under the chiral \( SU(3)_L \times SU(3)_R \) symmetry. 
Specifically, under a chiral transformation, $H_a \to H_b U^\dagger_{ba}$. Under heavy-quark spin symmetry, the transformation reads $H_a \to S H_a$, with \( S \) an element of the heavy-quark spin symmetry group. Finally, under a Lorentz transformation \( \Lambda \),
$H_a \to D(\Lambda) H_a D(\Lambda)^{-1}$, where \( D(\Lambda) \) denotes the spinor representation of the Lorentz group.

The effective Lagrangian for heavy mesons is constructed to respect both {chiral symmetry} and {heavy-quark symmetry}, and is typically written as
\begin{equation}
\begin{aligned}
\label{eq-couple}
\mathcal{L}_H 
&= - i\, \mathrm{Tr} \left( \bar{H}_a\, v_\mu \partial^\mu H_a \right)
+ \frac{i}{2}\, \mathrm{Tr} \left[ \bar{H}_a H_b\, v^\mu \left( U^\dagger \partial_\mu U + U \partial_\mu U^\dagger \right)_{ba} \right]  \\
&\quad + \frac{i g}{2}\, \mathrm{Tr} \left[ \bar{H}_a H_b\, \gamma_\nu \gamma_5 \left( U^\dagger \partial^\nu U - U \partial^\nu U^\dagger \right)_{ba} \right] 
+ \cdots,
\end{aligned}    
\end{equation}
where  the ellipsis denotes higher-order or symmetry-breaking contributions. 
The parameter \( g \) is the {axial coupling constant} of the heavy meson, while \( f_H \) denotes its {decay constant}.
The decay constant \( f_H \) is defined through the axial-vector current matrix element
$\langle 0 |\, \bar{q} \gamma^\mu \gamma_5 Q \,| H(p) \rangle = i\, f_H\, p^\mu,$
where \( Q \) and \( q \) represent the heavy and light quark fields, respectively, and \( p^\mu \) is the four-momentum of the heavy meson.

\subsection{Heavy Baryons as Bound States in the Skyrme Model}
Within the Skyrme model framework, a heavy baryon can be interpreted as a bound state of a heavy meson and a Skyrmion. In this picture, the Skyrmion is associated with the light-quark degrees of freedom, while the heavy meson encodes the dynamics of the heavy quark. Their interaction leads to the formation of a bound state that can be identified with a physical heavy baryon.

Starting from Eq.~(\ref{eq-couple}), when the chiral field is replaced by its classical Skyrmion configuration, the Lagrangian can be rewritten equivalently as 
\begin{equation}
\mathcal{L}
= -i\, \mathrm{Tr} \left( \bar{H}'_a\, v_\mu \partial^\mu H'_a \right)
+ \frac{i}{2}\, \mathrm{Tr} \left[ \bar{H}'_a H'_b\, v^\mu \left( \Sigma^\dagger \partial_\mu \Sigma \right)_{ba} \right]
+ \frac{i g}{2}\, \mathrm{Tr} \left[ \bar{H}'_a H'_b\, \gamma^\nu \gamma_5 \left( \Sigma^\dagger \partial_\nu \Sigma \right)_{ba} \right]
+ \cdots,
\end{equation}
where \( \Sigma(x) \) denotes the static Skyrmion field configuration, and \( H'_a \) represents the heavy meson field redefined in the Skyrmion background.
The interaction between the heavy meson and the Skyrmion then induces an effective binding potential that governs the dynamics of the heavy-baryon system.

Solving the corresponding bound-state equation yields the energy spectrum and wavefunctions of the heavy baryons. In the large-\(N_c\) limit, where the soliton mass scales as \( M_{} \sim N_c \) and the heavy-meson mass is treated as static \(( M_H \to \infty )\), the relative motion between the heavy meson and the Skyrmion can be well approximated by a harmonic-oscillator potential.
Within this approximation, the overlap wavefunction \( \phi(p) \) takes the form
\begin{equation}
\label{overlap}
    \phi (p) = \frac{1}{(\pi^2 \mu_b \kappa)^{3/4}} \exp(-p^2/(2 \sqrt{\mu_b \kappa})),
\end{equation}
where \( \kappa \) is defined as
$\kappa= \displaystyle g[ \frac{1}{3}[F'(0)]^3-\frac{5}{6} F'''(0)]$
with \( g \) denoting the axial coupling parameter, and \( \mu_b \) given by $\mu_b= (M_N M_B)/(M_N+M_B) $, with the $M_N$ and $M_B$ denoting masses of 
the nucleon and heavy meson, respectively.

\section{Weak Decays}

In this section, a general formulation for the study of purely baryonic weak decays of heavy baryons is presented. Specifically, to compute the decay rate of
$\Lambda_b\to p\bar{p}n$,
$\Gamma = \int d\Phi \overline{|\mathcal M|^2}$,
and the corresponding branching fraction, the analysis begins with the effective weak Hamiltonian. In the present case, the effective Hamiltonian can be written schematically as
\begin{equation}
    H_{\rm eff}
=\frac{G_F}{\sqrt2}V_{ub}V_{ud}^{}
\big(\bar u\gamma_\mu(1-\gamma_5)b\big)
\big(\bar d\gamma^\mu(1-\gamma_5)u\big)\,.
\end{equation}
Subsequently, the decay amplitude is given by
\begin{equation}
    \mathcal M
=\langle n(p_1,s_1),\bar p(p_2,s_2),p(p_3,s_3)|H_{\rm eff}|\Lambda_b(P,s)\rangle
=\frac{G_F}{\sqrt2}V_{ub}V_{ud}
\langle n\bar p p|J_\mu J'^{\mu}|\Lambda_b\rangle\,,
\end{equation}
where the currents are defined as
$J_\mu \equiv \bar u\gamma_\mu(1-\gamma_5)b$
and
$J'^{\mu}\equiv \bar d\gamma^\mu(1-\gamma_5)u.$

In the bound-state picture, the initial heavy baryons can be interpreted as
\begin{equation}
    |\Lambda_b(P,s)\rangle
=\sum_{s',s_B}\int d^3k
\phi(\mathbf k)
\mathcal{C}_{s',s_B;s}
|p(p',s')\rangle|B(p_B,s_B)\rangle,
\end{equation}
where $\mathcal{C}_{s',s_B;s}$ stands for the CG coefficiens, the kinematics are
$P^\mu=p'^{\mu}+p_B^\mu$ and $\mathbf k \equiv \mathbf p' - \mathbf p_B,
$
the bound-state wavefunction is given by Eq.~(\ref{overlap}).
Insert this expansion into the amplitude:
\begin{equation}
    \mathcal M
=\frac{G_F}{\sqrt2}V_{ub}V_{ud}
\sum_{s',s_B}\int d^3k
\phi(\mathbf k) \mathcal{C}_{s',s_B;s}
\langle n\bar p p|J_\mu J'^{\mu}|p(p',s')\rangle |B(p_B,s_B)\rangle.
\end{equation}

The decay amplitude receives contributions from several Feynman diagrams corresponding to different topologies. In the present study, attention is restricted to the dominant tree-level topology, shown on the left in Fig.~1, which captures the essential features of the baryonic weak decay process.
The nonfactorizable topology, shown on the right in Fig.~1, involves a multi-hadron matrix element and does not benefit from the color-singlet enhancement present in the factorizable contribution. Owing to its more intricate color structure and nonperturbative dynamics, and since the present work is intended as a first test of the Skyrme-model application, we assume this contribution to be suppressed at this stage.

\begin{figure}[htbp]
\label{fig}
  \centering
  \subfigure[]{
    \includegraphics[width=0.48\textwidth]{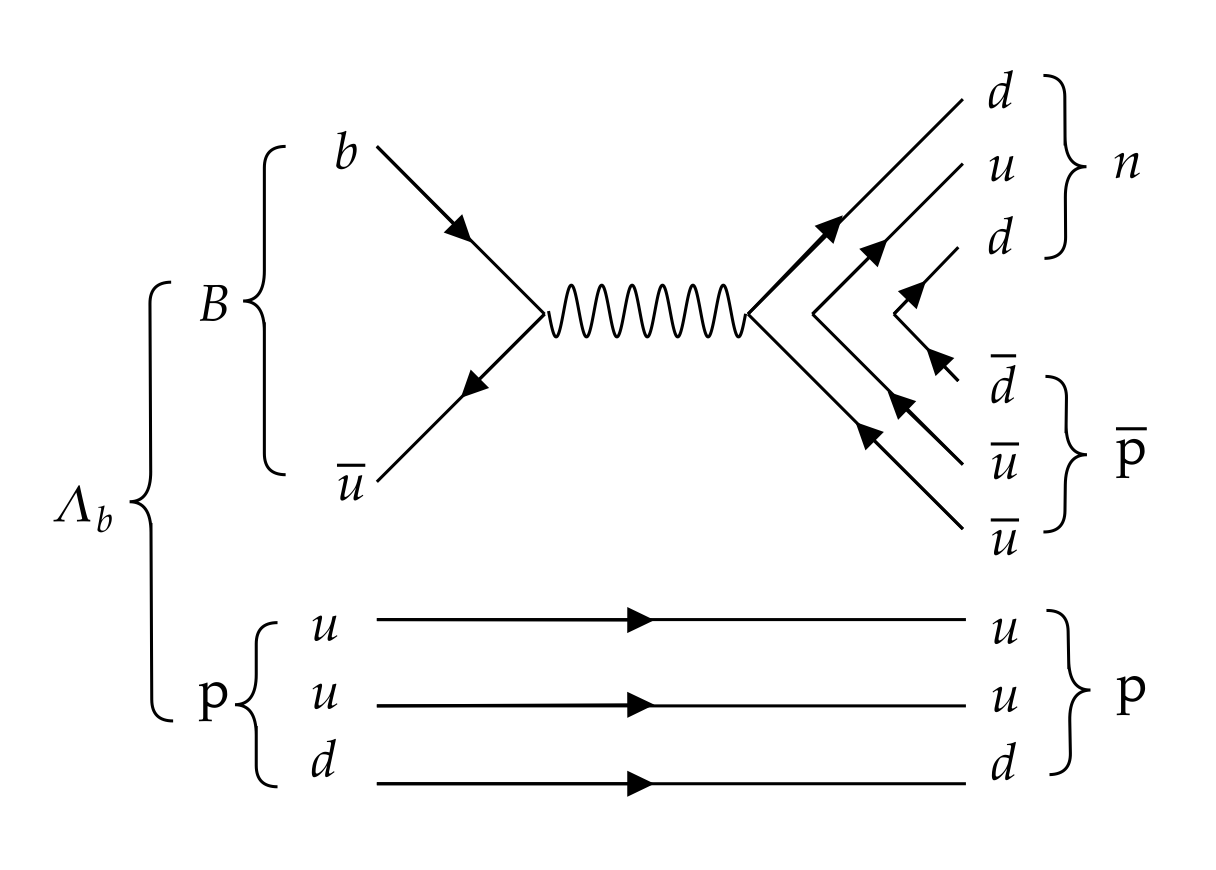}
    \label{fig:leading}
  }
  \hfill
  \subfigure[]{
    \includegraphics[width=0.48\textwidth]{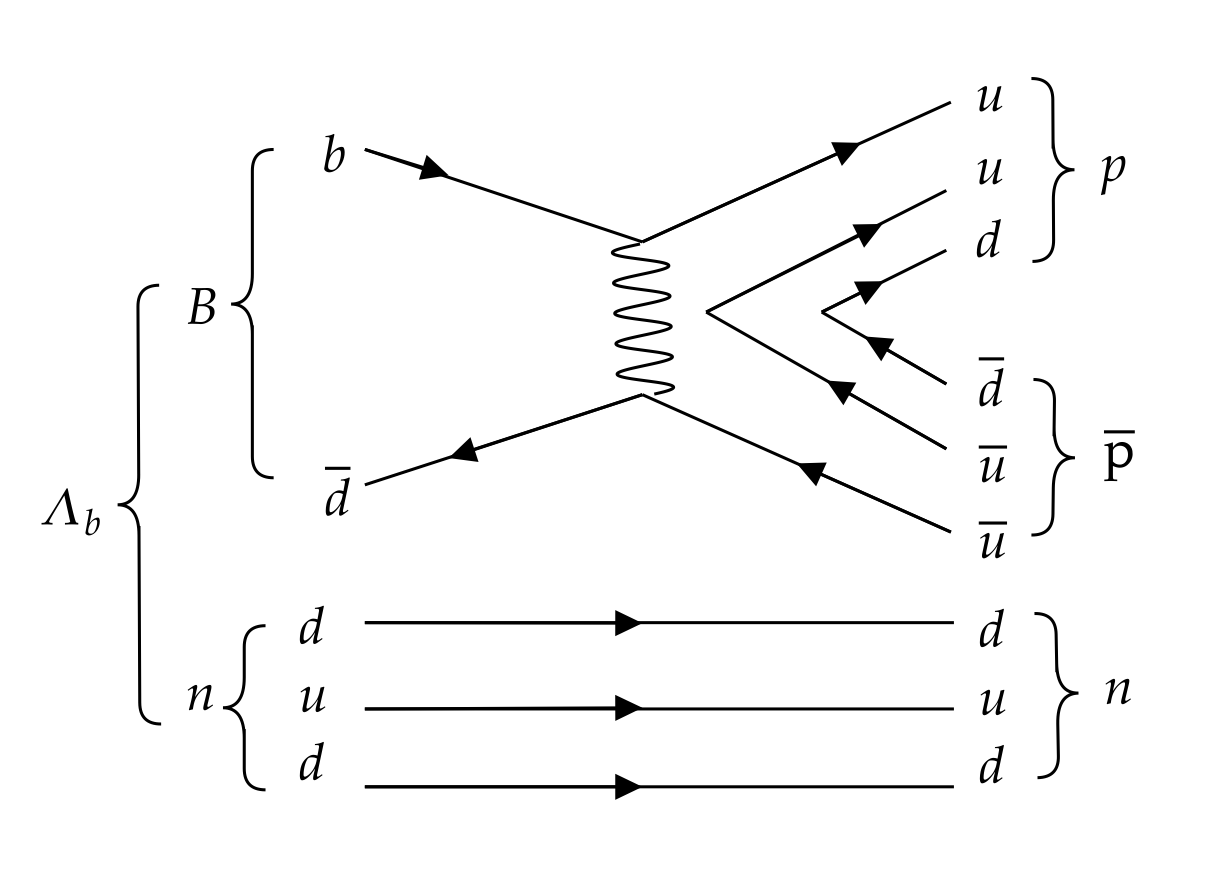}
    \label{fig:subleading}
  }
  \caption{Schematic diagrams for the tree-level contributions to the purely baryonic weak decay of $\Lambda_b\to p\bar{p}n$ within the Skyrme model framework.}
  \label{fig:decay-diagrams}
\end{figure}

The amplitude corresponding to the leading contribution implements the spectator approximation for the soliton sector and factorization for the heavy-meson component.
Explicitly, the spectator approximation implies that the proton in the initial bound state is identified with the final-state proton, $p(p',s') \to p(p_3,s_3)$, and the overlap produces a momentum–spin delta function.
Taking into account that the remaining matrix element can be further factorized, the full matrix element reduces to
\begin{equation}
    \langle n\bar p p|J_\mu J'^{\mu}|p B\rangle
\approx
\langle p(p_3,s_3)|p(p',s')\rangle
\langle n(p_1,s_1)\bar p(p_2,s_2)|J'^{\mu}|0\rangle
\langle 0|J_\mu|B(p_B,s_B)\rangle,
\end{equation}
indicating that the proton acts as a pure spectator, while the weak interaction operates solely on the heavy-meson component to produce the \( (\bar p, n) \) pair.
Consequently,
\begin{equation}
    \mathcal M
=\frac{G_F}{\sqrt2}V_{ub}V_{ud}
\sum_{s_B}\int d^3 k
\phi(\mathbf k)\mathcal{C}_{s',s_B;s}
\Big[ 2E_{p_3}\delta^{(3)}(\mathbf p_3-\mathbf p')\Big]
\langle n\bar p|J'^{\mu}|0\rangle
\langle 0|J_\mu|B\rangle\,,
\end{equation}
where $\langle 0|J_\mu|B\rangle$ can be expressed in terms of the heavy meson decay constant
$\langle 0|J_\mu|B(p_B)\rangle
= -i f_B q.$
The vacuum-to-\((n,\bar p)\) matrix element is expressed in terms of time-like form factors. A standard decomposition in the time-like region is given by
\begin{equation}
    \langle n\bar p|J'^{\mu}|0\rangle =
\bar u_n(p_1)\left[
F_1(q^2)\gamma^\mu+\frac{i}{2M_N}F_2(q^2)\sigma^{\mu\nu}q_\nu
-\left(G_A(q^2)\gamma^\mu+\frac{1}{2M_N}G_P(q^2)q^\mu\right)\gamma_5
\right]v_{\bar p}(p_2).
\end{equation}
Putting all ingredients together, the final expression for the decay amplitude can be written as
\begin{equation}
    \mathcal M=
\frac{G_F}{\sqrt2}V_{ub}V_{ud}
\Big[2E_{p_3}\phi(\mathbf k_\ast)\Big]
\mathcal{C}_{s',s_B;s}
\Big(-i f_B q_\mu\Big)
\bar u_n(p_1)\Gamma^\mu(q^2)v_{\bar p}(p_2),
\end{equation}
with
\begin{equation}
    \Gamma^\mu(q^2)=
F_1(q^2)\gamma^\mu+\frac{i}{2M_N}F_2(q^2)\sigma^{\mu\nu}q_\nu
-\left(G_A(q^2)\gamma^\mu+\frac{1}{2M_N}G_P(q^2)q^\mu\right)\gamma_5.
\end{equation}
The momentum transfer is defined as $q\equiv p_1+p_2$, and its relation to the final-state proton momentum $p_3$ is given by
\begin{equation}
    |\mathbf p_3(q^2)|  
=\frac{\lambda^{1/2}(M_{\Lambda_b}^2,q^2,M_N^2)}{2M_{\Lambda_b}},  
\quad  
E_3(q^2)=\frac{M_{\Lambda_b}^2+M_N^2-q^2}{2M_{\Lambda_b}},
\end{equation}
where the Källén function is defined by $\lambda(x,y,z)=x^2+y^2+z^2-2xy-2xz-2yz$.

After straightforward algebra, the squared amplitude can be expressed as
\begin{equation}
    \overline{|\mathcal M|^2}=
\left(\frac{G_F^2}{2}\right)
|V_{ub}V_{ud}|^2
\big(2E_{p_3}\big)^2|\phi(\mathbf k_\ast)|^2
f_B^2
q^2
\left|2M_N G_A(q^2)+\frac{q^2}{2M_N}G_P(q^2)\right|^2.
\end{equation}

Then the decay rate can be given
\begin{equation}
    \begin{aligned}
        \Gamma=&\int_{q^2_{\min}}^{q^2_{\max}} d q^2  
\frac{1}{2M_{\Lambda_b}}  
\overline{|\mathcal M(q^2)|^2}  
\frac{1}{256\pi^3}  
\frac{\lambda^{1/2}(M_{\Lambda_b}^2,q^2,M_N^2)}{M_{\Lambda_b}^2}  
\frac{\lambda^{1/2}(q^2,M_N^2,M_N^2)}{t}
\\
=&
\int_{q^2_{\min}}^{q^2_{\max}} d q^2 \frac{1}{512\pi^3 M_{\Lambda_b}^3 } 
\left(\frac{G_F^2}{2}\right)|V_{ub}V_{ud}|^2  
f_B^2  
|\phi(\mathbf k_\ast(q^2))|^2  
\big[2E_3(q^2)\big]^2  
\\ &
\times
\left|2M_N G_A(q^2)+\frac{q^2}{2M_N}G_P(q^2)\right|^2  
 \lambda^{1/2}(M_{\Lambda_b}^2,q^2,M_N^2) \lambda^{1/2}(q^2,M_N^2,M_N^2) \,,
    \end{aligned}
\end{equation}
where $\mathbf k_\ast$ is defined by
$|\mathbf k_\ast(q^2)|^2 =\displaystyle \frac{\lambda(M_{\Lambda_b}^2,q^2,M_N^2)}{M_{\Lambda_b}^2}$, and the kinematic limits are $q^2_{\min}=(2M_N)^2$
and
$q^2_{\max}=(M_{\Lambda_b}-M_N)^2.$

\section{Numerical Results}

To perform a quantitative evaluation of the decay width, the relevant input parameters are specified as follows~\cite{ParticleDataGroup:2024cfk}
\begin{align}
&
M_{\pi_0}=134.97~\text{MeV}, \,
M_{\pi_\pm}=139.57~\text{MeV}, \,
M_N = 938.3~\text{MeV},\,
M_{\Lambda_b} = 5.619~\text{GeV},
\nonumber\\&
M_{B^0}=5.279~\text{GeV},\,
M_{B^\pm}=5.279~\text{GeV},\,
\Gamma_{\text{total}}(\Lambda_b) = 4.48 (3) \times 10^{-13}~\text{GeV},
\nonumber\\&
F_\pi = 130.2(1.7)~\text{MeV},  \,
f_B = 190.0(1.3)~\text{MeV},\,
G_F=1.166\times 10^{-5}~\text{GeV}^{-2}, \,
\nonumber\\&
V_{ud}=0.97349(16), \,
V_{ub}=0.003732_{-0.000085}^{+0.000090}, \,
e = 4.84, \,
g=0.54(3), \nonumber
\end{align}
where $e$ is the Skyrme coupling constant and $g$ is the axial coupling constant.

\begin{figure}
    \centering
    \includegraphics[width=0.6\linewidth]{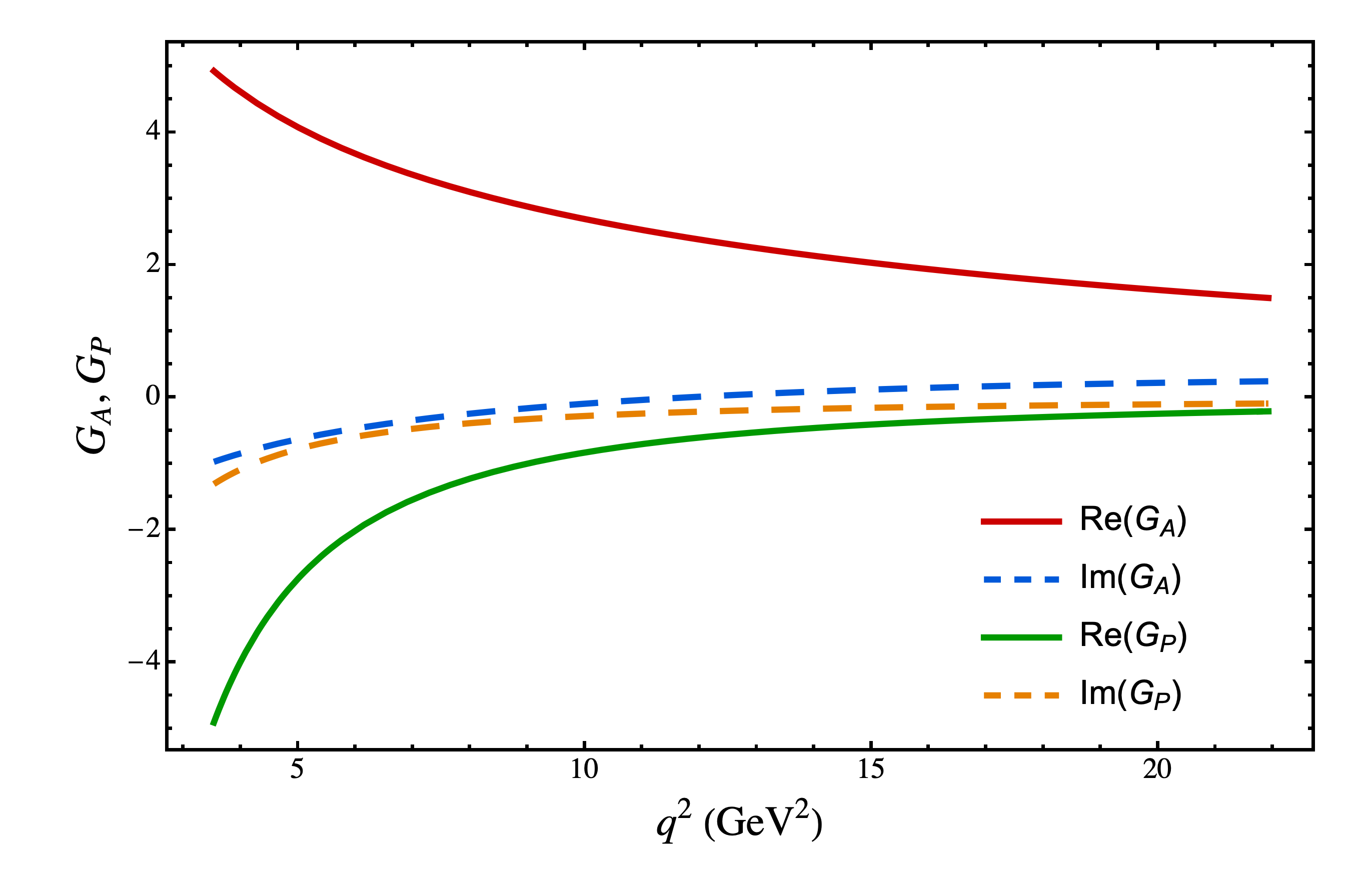}
    \caption{Real and imaginary parts of the $G_A$ and $G_P$ in the timelike region as functions of the momentum transfer squared $q^2$.}
    \label{fig:ff}
\end{figure}

Fig.~\ref{fig:ff} shows the axial-vector and pseudoscalar form factors \(G_A\) and \(G_P\) as functions of the momentum transfer squared \(q^2\) in the timelike region obtained in this work.
As shown in the figure, the real parts of \(G_A\) and \(G_P\) provide the dominant contributions in the considered \(q^2\) region, while the corresponding imaginary parts remain relatively small throughout the range studied here.
It should be noted that the present numerical results are still influenced by several computational approximations. In particular, the analytic continuation of the form factors is implemented using the Padé approximation, whose pole structure depends on the order of the approximation and may affect the behavior of the imaginary parts in the timelike region. In addition, only relatively low-order Padé approximants are adopted here for illustration, mainly to demonstrate the basic structure of the method and the numerical stability of the procedure.

Using these inputs, the decay width is evaluated numerically, leading to the corresponding branching fraction $\displaystyle\mathcal{B}(\Lambda_b \rightarrow p\,\bar{p}\,n)=(1.10 \pm0.27) \times 10^{-6}$.
Here the uncertainty only reflects the numerical inputs used in the calculation. Additional theoretical uncertainties from the large-\(N_c\) approximation ($1/N_c$) and the heavy-mass expansion ($1/M_B$) should also be understood, although they are not displayed explicitly for clarity.

\begin{figure}
    \centering
    \includegraphics[width=0.6\linewidth]{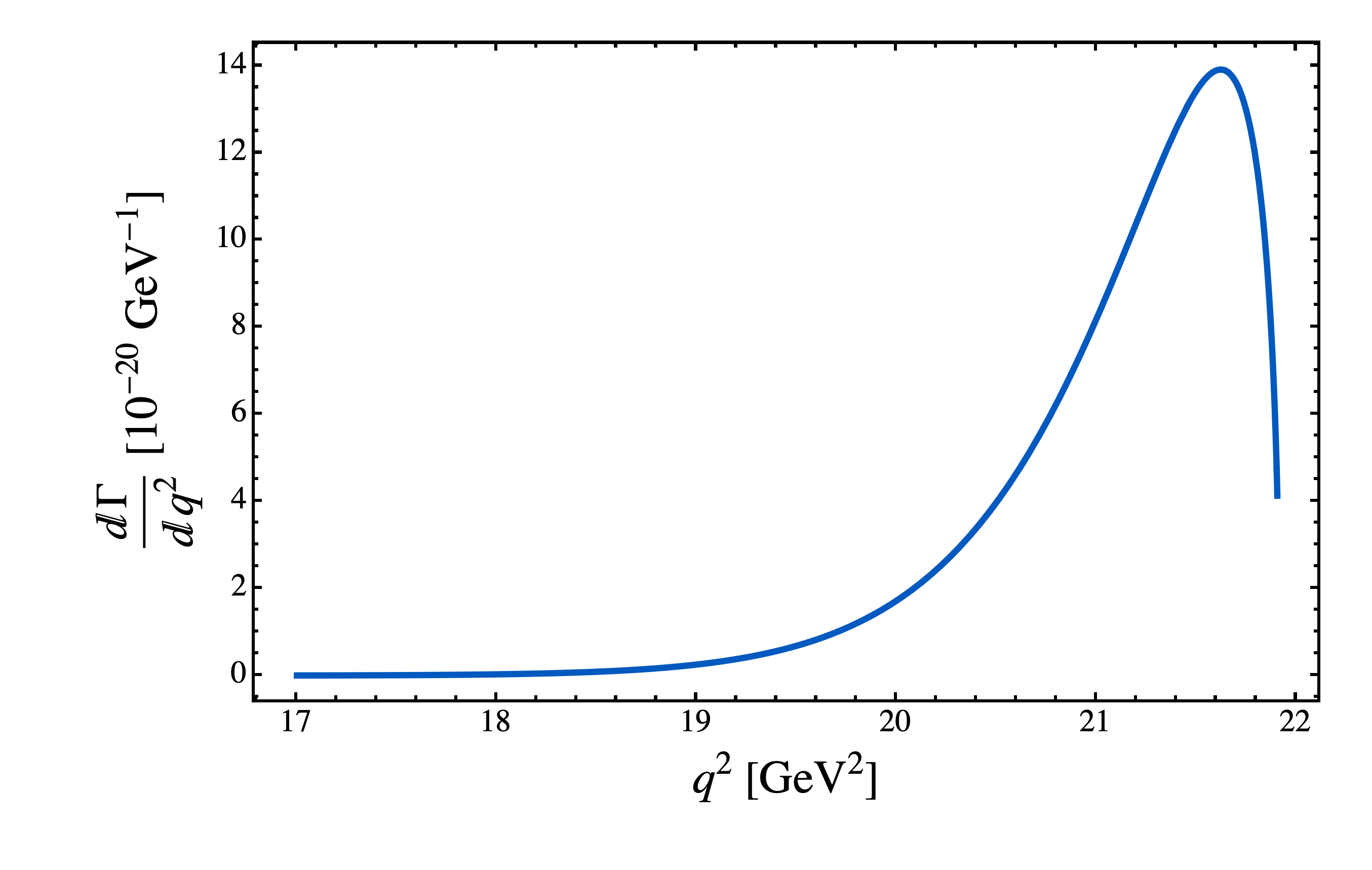}
    \caption{Differential decay rate as a function of the momentum transfer squared $q^2$.}
    \label{fig:decay}
\end{figure}

Fig.~\ref{fig:decay} illustrates the differential decay rate as a function of the momentum transfer squared \(q^2\), evaluated using the central values of the input parameters. As shown in the figure, the differential decay rate exhibits a enhancement near the kinematic endpoint. 
This sharp increase is mainly driven by the kinematic structure of the three-body phase space, together with the momentum dependence of the form factors in the timelike region.
In the present framework, the momentum transfer is largely determined by the transition of the heavy baryon into the \(p\bar p\) system. As a result, the available phase space is concentrated in the region of relatively large \(q^2\), which naturally enhances the differential decay rate near the upper end of the \(q^2\) spectrum.

A comparison with previous studies~\cite{Geng:2016drz,Hsiao:2018dqg}, which give a typical branching fraction $\mathcal{B} \sim 2 \times 10^{-6}$, is illuminating.
The result obtained here agrees with the previous estimates of $\mathcal{O}(10^{-6})$, but the center value is approximately half of the previous one.
This difference may suggest that additional mechanisms could play a role in the decay process. 
However, the present calculation should be regarded as a preliminary attempt, and a more refined analysis is required to clarify the origin of this discrepancy.

Following the same method, the present framework can be extended to other purely baryonic decays involving additional flavors, such as $\Lambda_b \to p\bar{p}\Lambda$, where the $\Lambda$ baryon is treated as a bound state of a kaon and a skyrmion~\cite{Callan:1985hy}. In particular, the heavier mass of the $\Lambda$ baryon reduces the available phase space and suppresses the decay rate. Then the branching fraction is expected to be smaller than the previous estimate of $\mathcal{B} \sim3\times10^{-6}$.

\section{conclusion}

In this paper, the purely baryonic weak decay of a heavy baryon is investigated within the Skyrme model. In this framework, the heavy baryon is treated as a bound state of a heavy meson and a soliton. In particular, the decay channel \( \Lambda_b \to p \,\bar p \, n \) is studied in detail.
Employing the large-\(N_c\) approximation, the leading contribution to the decay amplitude arises from the spectator mechanism, under which the relevant hadronic matrix elements factorize into a set of tractable components.
The corresponding matrix elements are derived, leading to an explicit prediction for the branching fraction. 
It turns out that our result is approximately half of the previous estimate, which calls for a more detailed investigation in future studies.  
The present calculation provides only a rough estimate, as it is restricted to the leading contribution in the large-\(N_c\) expansion, and relativistic effects such as those discussed in Ref.~\cite{Ji:1991ff} are not included. These ingredients may influence the soft and near-threshold dynamics.  
The present framework can also be extended to include additional flavors and to study other purely baryonic decay modes such as $\Lambda_b \to p\bar{p}\Lambda$.
These may provide useful guidance for future theoretical studies and experimental searches for purely baryonic decays.

\section*{Acknowledgments} 
This work is supported in part by 
the National Natural Science Foundation of China (NSFC) under Grant No. 12547104.


\bibliography{main}

@article{Geng:2016drz,
    author = "Geng, C. Q. and Hsiao, Y. K. and Rodrigues, Eduardo",
    title = "{Exploring the simplest purely baryonic decay processes}",
    eprint = "1603.05602",
    archivePrefix = "arXiv",
    primaryClass = "hep-ph",
    doi = "10.1103/PhysRevD.94.014027",
    journal = "Phys. Rev. D",
    volume = "94",
    number = "1",
    pages = "014027",
    year = "2016"
}

@article{Hsiao:2018dqg,
    author = "Hsiao, Y. K. and Geng, C. Q. and Rodrigues, Eduardo",
    title = "{Purely baryonic decay processes}",
    eprint = "1806.00861",
    archivePrefix = "arXiv",
    primaryClass = "hep-ph",
    doi = "10.1038/s41598-018-37743-9",
    journal = "Sci. Rep.",
    volume = "9",
    pages = "1358",
    year = "2019"
}

@article{Weinberg:1978kz,
    author = "Weinberg, Steven",
    editor = "Deser, S.",
    title = "{Phenomenological Lagrangians}",
    reportNumber = "HUTP-78-A051A",
    doi = "10.1016/0378-4371(79)90223-1",
    journal = "Physica A",
    volume = "96",
    number = "1-2",
    pages = "327--340",
    year = "1979"
}

@article{Weinberg:1990rz,
    author = "Weinberg, Steven",
    title = "{Nuclear forces from chiral Lagrangians}",
    reportNumber = "UTTG-31-90",
    doi = "10.1016/0370-2693(90)90938-3",
    journal = "Phys. Lett. B",
    volume = "251",
    pages = "288--292",
    year = "1990"
}

@article{Scherer:2002tk,
    author = "Scherer, Stefan",
    editor = "Negele, John W. and Vogt, E. W.",
    title = "{Introduction to chiral perturbation theory}",
    eprint = "hep-ph/0210398",
    archivePrefix = "arXiv",
    reportNumber = "MKPH-T-02-09",
    journal = "Adv. Nucl. Phys.",
    volume = "27",
    pages = "277",
    year = "2003"
}

@article{Skyrme:1961vq,
    author = "Skyrme, T. H. R.",
    title = "{A Nonlinear field theory}",
    doi = "10.1098/rspa.1961.0018",
    journal = "Proc. Roy. Soc. Lond. A",
    volume = "260",
    pages = "127--138",
    year = "1961"
}

@article{Skyrme:1962vh,
    author = "Skyrme, T. H. R.",
    title = "{A Unified Field Theory of Mesons and Baryons}",
    doi = "10.1016/0029-5582(62)90775-7",
    journal = "Nucl. Phys.",
    volume = "31",
    pages = "556--569",
    year = "1962"
}

@article{Adkins:1983ya,
    author = "Adkins, Gregory S. and Nappi, Chiara R. and Witten, Edward",
    title = "{Static Properties of Nucleons in the Skyrme Model}",
    reportNumber = "PRINT-83-0493 (IAS,PRINCETON)",
    doi = "10.1016/0550-3213(83)90559-X",
    journal = "Nucl. Phys. B",
    volume = "228",
    pages = "552",
    year = "1983"
}

@article{Ma:2016npf,
    author = "Ma, Yong-Liang and Harada, Masayasu",
    title = "{Lecture notes on the Skyrme model}",
    eprint = "1604.04850",
    archivePrefix = "arXiv",
    primaryClass = "hep-ph",
    month = "4",
    year = "2016"
}

@article{Braaten:1986md,
    author = "Braaten, Eric and Tse, Sze-Man and Willcox, Charles",
    title = "{Electroweak Form-factors of the Skyrmion}",
    reportNumber = "ANL-HEP-PR-86-22",
    doi = "10.1103/PhysRevD.34.1482",
    journal = "Phys. Rev. D",
    volume = "34",
    pages = "1482",
    year = "1986"
}

@article{Alberto:2015xea,
    author = "Alberto, Pedro and Drago, Alessandro and Mangoni, Alessio and Moretti, Simone and Pacetti, Simone",
    title = "{Analytic continuation of nucleon electromagnetic form factors in the time-like region}",
    eprint = "1507.05425",
    archivePrefix = "arXiv",
    primaryClass = "hep-ph",
    doi = "10.1088/1361-6471/abfbc2",
    journal = "J. Phys. G",
    volume = "48",
    number = "8",
    pages = "085007",
    year = "2021"
}

@article{Wise:1992hn,
    author = "Wise, Mark B.",
    title = "{Chiral perturbation theory for hadrons containing a heavy quark}",
    reportNumber = "CALT-68-1765",
    doi = "10.1103/PhysRevD.45.R2188",
    journal = "Phys. Rev. D",
    volume = "45",
    number = "7",
    pages = "R2188",
    year = "1992"
}

@book{Manohar:2000dt,
    author = "Manohar, Aneesh V. and Wise, Mark B.",
    title = "{Heavy quark physics}",
    doi = "10.1017/9781009402125",
    isbn = "978-0-521-03757-0, 978-1-009-40212-5",
    volume = "10",
    year = "2000"
}

@article{Jenkins:1992se,
    author = "Jenkins, Elizabeth Ellen and Manohar, Aneesh V. and Wise, Mark B.",
    title = "{The Baryon Isgur-Wise function in the large N(c) limit}",
    eprint = "hep-ph/9208248",
    archivePrefix = "arXiv",
    reportNumber = "UCSD-PTH-92-27, CALT-68-1809",
    doi = "10.1016/0550-3213(93)90257-P",
    journal = "Nucl. Phys. B",
    volume = "396",
    pages = "38--52",
    year = "1993"
}

@article{Jenkins:1992zx,
    author = "Jenkins, Elizabeth Ellen and Manohar, Aneesh V. and Wise, Mark B.",
    title = "{Baryons containing a heavy quark as solitons}",
    eprint = "hep-ph/9205243",
    archivePrefix = "arXiv",
    reportNumber = "CALT-68-1783-REV, UCSD-PTH-92-17-REV",
    doi = "10.1016/0550-3213(93)90256-O",
    journal = "Nucl. Phys. B",
    volume = "396",
    pages = "27--37",
    year = "1993"
}

@book{Weigel:2008zz,
    author = "Weigel, Herbert",
    title = "{Chiral Soliton Models for Baryons}",
    volume = "743",
    year = "2008"
}

@article{Holzwarth:1985rb,
    author = "Holzwarth, G. and Schwesinger, B.",
    title = "{Baryons in the Skyrme Model}",
    reportNumber = "PRINT-86-0159 (SIEGEN)",
    doi = "10.1088/0034-4885/49/8/001",
    journal = "Rept. Prog. Phys.",
    volume = "49",
    pages = "825",
    year = "1986"
}

@article{ParticleDataGroup:2024cfk,
    author = "Navas, S. and others",
    collaboration = "Particle Data Group",
    title = "{Review of particle physics}",
    doi = "10.1103/PhysRevD.110.030001",
    journal = "Phys. Rev. D",
    volume = "110",
    number = "3",
    pages = "030001",
    year = "2024"
}

@article{Callan:1985hy,
    author = "Callan, Jr., Curtis G. and Klebanov, Igor R.",
    title = "{Bound State Approach to Strangeness in the Skyrme Model}",
    reportNumber = "Print-85-0733 (PRINCETON)",
    doi = "10.1016/0550-3213(85)90292-5",
    journal = "Nucl. Phys. B",
    volume = "262",
    pages = "365--382",
    year = "1985"
}

@article{Ji:1991ff,
    author = "Ji, Xiang-Dong",
    title = "{A Relativistic skyrmion and its form-factors}",
    doi = "10.1016/0370-2693(91)91185-X",
    journal = "Phys. Lett. B",
    volume = "254",
    pages = "456--461",
    year = "1991"
}

\end{document}